\documentstyle[pptwocol,aaspp4,astrobib]{preprint}

\newcommand{\ltappeq}{\raisebox{-0.6ex}{$\,\stackrel
{\raisebox{-.2ex}{$\textstyle <$}}{\sim}\,$}}
\newcommand{\gtappeq}{\raisebox{-0.6ex}{$\,\stackrel
{\raisebox{-.2ex}{$\textstyle >$}}{\sim}\,$}}

\begin{document}

\title{\bf{Recovery of 29-s Oscillations in HST/FOS Eclipse
Observations of the Cataclysmic Variable UX Ursae Majoris\footnote{Based on 
observations with the NASA/ESA {\em Hubble Space Telescope}, 
obtained at the Space Telescope Science Institute, which is 
operated by the Association of Universities for Research in 
Astronomy, Inc., under NASA contract NAS~5-2655}}}

\author{Christian Knigge}
\affil{Space Telescope Science Institute, 3700 San Martin Drive,
Baltimore, MD 21218 \\ knigge@stsci.edu} 

\author{Nick Drake}
\affil{The University of Southampton, Department of Physics \&
Astronomy, Southampton~SO17~1BJ, United Kingdom \\
nsd@astro.soton.ac.uk}

\author{Knox S. Long}
\affil{Space Telescope Science Institute, 3700 San Martin Drive,
Baltimore, MD 21218 \\ long@stsci.edu} 

\author{Richard A. Wade} 
\affil{The Pennsylvania State University, Department of Astronomy and
Astrophysics, \\ 525 Davey Laboratory, University Park, PA 16802 \\
wade@astro.psu.edu}

\author{Keith Horne}
\affil{Department of Physics and Astronomy, The University of
St. Andrews, North Haugh, St. Andrews, Fife, KY16 9SS, UK \\
kdh1@st-andrews.ac.uk}

\and

\author{Raymundo Baptista} 
\affil{Departamento de Fisica, Universidade
Federal de Santa Catarina, Campus Universitario, Trindade, 88040
Florianopolis, Brasil \\
bap@fsc.ufsc.br}

\begin{abstract}

Low amplitude ($\simeq 0.5\%$) 29-s oscillations have been detected in
HST/FOS eclipse observations of the 
nova-like cataclysmic variable UX~UMa. These are the same dwarf nova-type 
oscillations that were originally discovered in this system by Warner 
\& Nather in 1972. The 29-s oscillations are seen in one 
pair of eclipse sequences obtained with the FOS/PRISM in November of
1994, but not in a similar pair obtained with the FOS/G160L grating in
August of the same year. The oscillations in the PRISM data are 
sinusoidal to within the small observational errors and undergo 
an approximately $-360^o$ phase shift
during eclipses (i.e. one cycle is lost). Their amplitudes are highest
at pre-eclipse orbital phases and exhibit a rather gradual eclipse
whose shape is roughly similar to, though perhaps slightly narrower
than, UX~UMa's overall light curve in the PRISM bandpass
(2000~\AA~-~8000~\AA). 

Spectra of the oscillations have
been constructed from pre-, mid- and post-eclipse data segments of 
the November observations. The spectra obtained from the
out-of-eclipse segments are extremely blue, and only lower limits 
can be placed on the temperature of the source which dominates the 
modulated flux at these orbital phases. Lower limits derived from
blackbody (stellar atmosphere) model fits to these data are 
$\geq$~95,000~K ($\geq$~85,000~K); the corresponding 
upper limits on the projected area of this source are all $< 2\%$ 
of the WD surface area. By contrast, oscillation spectra derived 
from mid-eclipse data segments are much redder. Fits to these 
spectra yield temperature estimates in the range $20,000$~K$\ltappeq T
\ltappeq 30,000$~K for both BB and SA models and corresponding
projected areas of a few percent of the WD surface area. These
estimates are subject to revision if the modulated emission is
optically thin.

We suggest that the ultimate source of the oscillations is a hot,
compact region near disk center, but that a significant fraction of
the observed, modulated flux is due to reprocessing of the light
emitted by this source in the accretion disk atmosphere. The 
compact source is occulted at orbital phases near mid-eclipse, 
leaving only part of the more extended reprocessing 
region(s) to produce the weak oscillations that persist even at
conjunction.

The highly sinusoidal oscillation pulse shape does not permit the
identification of the compact component in this model 
with emission produced by a rotating disturbance in the inner
disk or in a classical, equatorial boundary layer. Instead, this
component could arise in a bright spot on the surface of the WD,
possibly associated with a magnetic pole. However, a standard 
intermediate polar model can also be ruled out, since UX~UMa's
oscillation period has been seen to change on time-scales much shorter
than the minimum time-scale required to spin up the WD by accretion
torques. A model invoking magnetically controlled accretion onto {\em
differentially rotating} WD surface layers may be viable, but needs
more theoretical work.

\end{abstract}


\section{Introduction}
\label{introduction}

Cataclysmic variable stars (CVs) are semi-detached
binary systems in which mass is transferred from a Roche 
lobe-filling, approximately main-sequence secondary onto an 
accretion disk around a mass-gaining white dwarf (WD) primary. 
In the nova-like (NL) CVs, the accretion disk is optically thick 
and dominates the ultraviolet (UV) and optical light. This makes NL
CVs excellent targets for observational studies of accretion disk
physics, particularly since their binary nature often allows
their system parameters to be determined quite accurately, 

In a previous paper (\citeNP{me7}, hereafter Paper~I), we presented a   
set of spectrally- and orbital phase-resolved eclipse observations of
the NLCV UX~UMa that were obtained with the {\em Faint Object
Spectrograph} (FOS) onboard the {\em Hubble Space Telescope} (HST). In
our analysis of the data in Paper~I, the emphasis was on the spectral
characteristics of the system components that could be isolated in the
data (the accretion disk, the bright spot and the uneclipsed
light). Here, we focus on UX~UMa's 29-s 
oscillations which exist in some of our data. These 
oscillations were originally discovered in this system by Warner 
\& Nather (1972, hereafter WN72) and belong to the class of 
``dwarf nova oscillations'' (DNOs). The name derives from
the fact that DNOs -- small-amplitude ($\ltappeq 0.5$\%), short-period
($7$~s$ \ltappeq P_{DNO} \ltappeq 40$~s) and reasonably coherent ($10^4
\ltappeq Q=| \dot{P_{DNO}} |^{-1} \ltappeq 10^6$) oscillations -- have
been observed mainly in dwarf novae (DNe) in outburst, although a few
other NLs also exhibit such modulations (e.g. \citeNP{warner6}). Most
observations of DNOs have been at optical wavelengths, but some
systems are known to show periodic x-ray oscillations as well. The 
very stable ($Q > 10^{12}$) oscillations displayed by two intermediate
polars (IPs) -- AE~Aqr and DQ~Her -- may also be related to the DNO
phenomenon.\nocite{warner4,nather1} 

The origin of DNOs is not certain. Two relatively plausible scenarios 
invoke magnetically controlled accretion close to the WD, and 
relatively long-lived disturbances in the inner disk or the boundary 
layer (see Warner~1995 for a recent review). In both models, the
ultimate source of the oscillations is near the center of the 
accretion disk and the period of the oscillations is identified with 
the dynamical (rotational) time scale in the vicinity of the
source. In fact, the latter identification is common to essentially
all models for the origin of DNOs, mainly because all other 
time scales (e.g. thermal and viscous ones) are long compared to
typical DNO periods (tens of seconds). Thus DNOs may be 
a unique probe of the physical conditions very close to the
central source in those disk-accreting systems that exhibit these
oscillations. Moreover, given the relative ubiquity of DNOs among CVs,
their excitation mechanism is likely to be of quite general
significance to our understanding of accretion disk physics.

The plan of this paper is as follows: in Section~\ref{observations},
we briefly describe the relevant aspects of the observations and their
reduction. UX~UMa's 29-s oscillations are then recovered and analyzed
in Section~\ref{oscillations}. There we consider the eclipse behavior,
mean pulse shape and coherence properties, as well as the spectrum of
the oscillations. We discuss our results in Section~\ref{discussion},
focusing particularly on the impact of our new data on models for the
origin of DNOs. Finally, we present our conclusions in
Section~\ref{conclusions}.

\section{Observations}
\label{observations}

UX~UMa was observed with the FOS on HST in August and November of
1994. Two eclipses were followed in each epoch. The two August eclipse
observations (Runs~1 and 2) were carried out with the G160L grating 
and covered consecutive orbital cycles; the two November observations 
(Runs~3 and 4) were obtained with the PRISM and were separated by two
unobserved eclipses. All observations were carried out in RAPID mode,
with a new exposure beginning every 5.4~s. 

The nominal wavelength coverages of the spectral elements used in 
the two observing epochs were 1140~\AA~-~2508~\AA~(G160L; August) 
and 1850~\AA~-~8950~\AA~(PRISM; November). However, for practical 
purposes (see Paper~I), we restricted our analysis of the data to 
1230~\AA~-~2300~\AA~for G160L spectra and 2000~\AA~-~8000~\AA~for
PRISM spectra. The spectral resolution of the G160L was 6.6~\AA~FWHM
with our instrumental set-up; that of the PRISM varied from a few 
Angstrom near the short wavelength end to a few hundred Angstrom near
the long wavelength end. In observations with the G160L grating, 
the zeroth order light is also recorded and can be used to construct a
broad-band optical/UV light curve. The zeroth order light has a
bandpass with full-width at half-response of 1900~\AA~and a pivot
wavelength of 3400~\AA. For more details on the observations and their
reduction, the reader is referred to Paper~I.

In the top panels of Figure~\ref{fig-lightpower}, ``white-light'' 
light curves are shown for the four observing runs. These have been 
constructed as time-series of the average flux across the full adopted
wavelength range of the relevant dispersion element in each exposure. For the 
G160L observing sequences, the zeroth order time series are also shown. 
Note that UX~UMa was~$\sim$50\% brighter in November than in August,
which, if due to a change in the accretion rate, indicates an ncrease
in $\dot{M}_{acc}$ by $\gtappeq 50\%$ (Paper~I). 

Lomb-Scargle power spectra \cite{lomb1,scargle1} calculated from 
these light curves are plotted in the bottom 
panels of Figure~\ref{fig-lightpower}. Flickering -- the non-periodic,
intrinsic variability common to essentially all types of CVs -- shows
up in these power spectra as a general increase in power towards low
frequencies. Peaks at frequencies corresponding to periods of about
29~s are nevertheless easily identified in the power spectra of Runs~3
and 4 (November; PRISM) since they are well separated from the
low-frequency flickering-dominated regime. Similar peaks are not seen
in any of the power spectra constructed for Runs~1 and 2 (August;
G160L).

\section{Recovery and Analysis of the 29-second Oscillations}
\label{oscillations}

Before embarking on the analysis of the oscillations in our own data,
we briefly review the more striking properties of the 29-s
oscillations observed by WN72 and NR74. The oscillations analyzed by 
WN72 had an amplitude of about 0.002 mag (approximately 0.2\%) in
white optical light,\footnote{Throughout this paper, we will use the
terms ``peak-to-peak amplitude'' and ``amplitude'' to denote 
the full peak-to-peak amplitude of a periodic 
signal and one half thereof, respectively.} 
were detected at orbital phases before, during and after
eclipse, and had a pulse shape that was sinusoidal to within
observational uncertainties. In addition, WN72 found that the
oscillations were not present in all of their observing runs, and that
their period was slightly, but measurably different in the two runs in
which they were detected. Finally, NR74 showed that the
oscillations underwent a -360$^o$ phase shift (meaning
that one oscillation cycle was lost) during eclipse, even though no
obvious eclipse-related amplitude modulations were detected.

\subsection{Wavelength-averaged properties}
\label{waveave}

Given this background, we decided to demodulate our time series using
the ``sliding sine fit'' technique described by NR74. The first
step in this procedure is to apply a high-pass filter to
the light curves to remove long term trends. We filtered the data by 
subtracting a 5-point running mean from each
datum in each time series. Given our time resolution of 5.4~s, a
5-point boxcar smoothing corresponds to averaging over about 
one 29-s oscillation cycle. We also experimented with an 11-point 
boxcar filter (corresponding to an average over 2 oscillation cycles)
and found that this gave essentially identical, but somewhat noisier 
results. The resulting high-pass filtered time-series are shown in
Figures~\ref{fig-cohere3} and~\ref{fig-cohere4}. The oscillations are
seen clearly in the filtered data and appear to be approximately
sinusoidal. In addition, there seems to be an eclipse-related
decrease in the oscillation amplitudes. 

Next, power spectra were calculated from the filtered data
sets in order to identify the oscillation periods more accurately. 
These power spectra are shown in Figure~\ref{fig-power_30s}, treating 
pre-, mid- and post-eclipse phases separately to avoid averaging over
possible eclipse phase 
shifts. The appropriate periods to be used in demodulating the data
are easily identified from the power spectra -- they are the
pre-eclipse periods of 28.77 s (Run~3) and 28.60 s (Run~4) -- but a 
number of other points are also worth noting from this figure.
First, the peak powers in all six power spectra are highly
significant. This is immediately obvious from the high power levels in
the Lomb-Scargle periodograms (in which power is normalized to the
variance of the data) and supported by robust, distribution-free 
randomization tests we have carried out. Thus the 29-s oscillations 
are detected at all orbital phases, consistent with the results of
WN72. Second, the peaks are highest in the power spectra calculated
from the pre-eclipse data segments, suggesting that the amplitude of the
oscillations was also highest there. This again gives hope that we may
be able to detect and characterize amplitude variations as a function of
orbital phase for the first time. Third, the power spectra suggest
that the oscillation period at mid-eclipse orbital phases is slightly, but
systematically lower than at pre- and post-eclipse phases. This is the
first hint that we are seeing the eclipse phase shift described by
NR74 in our data as well. (Note that if a time series is demodulated
by fitting sinusoids with fixed period but variable phase, a linear 
phase shift indicates that the signal in the data repeats on a 
slightly different period than assumed in the fit.) Fourth, the
periods detected in the pre-, mid- and post-eclipse time series are
equal for both November observing runs, to within the uncertainties
defined by the widths of the peaks in the power spectra. Fifth, the 
inset in Figure~\ref{fig-power_30s} shows that we do not detect any
signal at the first overtone frequency in the power spectrum of the
data segment in which the oscillations are detected most cleanly
(Run~3 pre-eclipse). This gives quantitative support to the visual
impression from Figures~\ref{fig-cohere3}~and~\ref{fig-cohere4} 
that the oscillations are sinusoidal to high accuracy. To test for the
presence of sub-harmonics, we have also inspected the power spectrum
constructed from the Run~3 pre-eclipse light curve after filtering the
data with several wider filters, ranging in width from  2 to 4
oscillation cycles. No power excesses were found at frequencies
corresponding to integral multiples of the 29-s period.

The highly sinusoidal character of the oscillations is illustrated
more directly in Figure~\ref{fig-sinefit}. This shows the result of
folding the Run~3 pre-eclipse, high pass-filtered time series onto the
28.77~s oscillation period and binning the resulting light curve into
20 non-overlapping phase bins. The figure establishes that the
mean pulse shape is a simple sinusoid to within the small 
observational errors. This result is not an artifact of the narrow 
filter width used: a fixed-period sinusoid fit to the same data
segment after applying a filter of twice the original width still
gives an excellent reduced $\chi^{2}$ of 0.8. Note that each 
vertical error bar in Figure~\ref{fig-sinefit} gives the error on 
the mean of all the points in that phase bin; the standard deviation
of the points in each 
bin, which measures the spread of the data points around the mean, is 
larger (by definition) and plotted on the bottom axis. We emphasize 
this distinction because the small error bars in 
Figure~\ref{fig-sinefit} imply only that there was a well-defined {\em
mean} pulse profile during the relevant pre-eclipse time interval, but
do not rule out the possibility that significant amplitude 
fluctuations may also have occurred. Nevertheless, the combination of 
a highly sinusoidal mean pulse shape and the total absence of power at
the first overtone frequency does allow us to discard models that
predict more complex light modulations (see
Section~\ref{discussion}).

With the filtered data sets and good estimates of the 
pre-eclipse oscillation periods in hand, we proceeded to fit 
sinusoids of fixed period (28.77 s for Run~3; 28.60s for Run~4) 
to blocks of data containing 16 points (3 oscillation cycles)
each. Following NR74, we allowed approximately 50\% overlap between 
successive data segments being fit. The results of demodulating 
both PRISM time series in this way are shown in
Figure~\ref{fig-thirty}. In the top panels of this figure, we plot 
again the white-light light curves themselves, with insets showing
small sections of the light curves complete with error bars. Note that
the 29-s oscillations can actually be seen directly even in the
unfiltered data. The lower panels show the filtered light curves, the
oscillation amplitudes and the oscillation phases determined from the sinusoid
fits. The filtered light curves and oscillation amplitudes are shown in
both absolute (ergs cm$^{-2}$ s$^{-1}$ \AA$^{-1}$) and relative
(percentage of total light at that phase) units in order to test if
and how the oscillation amplitudes are correlated with the total flux
at a given orbital phase.

Perhaps the most important new result from Figure~\ref{fig-thirty} is
that orbital phase-linked amplitude variations are clearly detected
for the first time in UX~UMa. The absolute
amplitudes are greatest at pre-eclipse orbital phases, decline during
eclipse and recover only slightly at post-eclipse phases. When
expressed as a fraction of the total light at a given orbital phase,
the oscillation amplitudes (which are about 0.5\% at pre-eclipse 
phases) exhibit only a weak eclipse feature, but show some spikes just
before and after eclipse. This indicates that the oscillation eclipse 
is similar in relative depth to that of the total light, but has a 
slightly narrower width. 

Our detection of oscillation eclipses contrasts with the apparent
absence of orbital phase-related amplitude modulations in the
observations analyzed by NR74. Since their data were of much lower
quality, it is not clear how much weight should be attached to this
difference. NR74 estimate that amplitude changes greater than a factor
of two  would probably have been detectable in their data -- a criterion
satisfied by the eclipses of the oscillations in our observations. 
However, to combat noise, NR74 found it necessary to average over 
several eclipses in their search for amplitude modulations, so 
their non-detection might have been caused by secular and/or 
stochastic variability between runs. The amplitudes of the
oscillations in our data certainly do exhibit sizeable fluctuations
even away from eclipse. More high-quality data will be required to
decide whether UX~UMa's 29-s oscillations always undergo eclipses 
or not.

The eclipse-related phase shift noted by NR74 is easily seen in 
Figure~\ref{fig-thirty}. The absolute value of the total shift
over the course of the eclipse actually appears to be somewhat larger
than 360$^o$ in our data. However, it is possible that the
post-eclipse oscillations once again attained more accurate
coherence with their pre-eclipse counterparts at orbital phases 
beyond the end of the PRISM observing sequences. For Run~3, at least,
Figure~\ref{fig-thirty} also hints at the phase shift progressing in 
three stages. First, there 
appears to be a relatively steep, approximately linear decline. Next,
there is a short interval just after mid-eclipse in which there is
essentially no further phase drift. Following this, the shift is
completed in roughly the same manner as in the first stage. However,
given the limited amount of data on which it is based, this
description is probably not unique. NR74 also saw a departure from
linearity in the phase shifts they detected, which they described
as a BS-related ``lump''.

Since different periods were used in the sinusoid fits to the 
Run~3 and~4 data, the oscillation phases predicted by the fits cannot
be compared between runs. In Figure~\ref{fig-thirty}, we have
therefore shifted the oscillation phases derived from both runs to a
common mean pre-eclipse value of 180$^o$. It would nevertheless 
be useful to determine whether the oscillations in both runs could
have been coherent with each other if the true (out-of-eclipse) 
oscillation period is assumed to have remained constant. In that 
case, the difference between the values measured from the relevant
power spectra ($\Delta P = P_{Run~3} - P_{Run~4} = 0.17$~s) would have
to be attributed to observational error. Adopting $P = 28.7$~s as an
estimate of the oscillation period and noting that the time difference
between the end of Run~3 and the start of Run~4 is approximately
$\Delta t \simeq 47,000$~s, we find that the uncertainty in the cycle 
count would be about $\Delta E = (\Delta t \Delta P) / (P^2)
\simeq 10 $~cycles at the start of Run~4. Thus the data cannot be
used to test the coherence properties of the oscillations on time
scales as long as the gap between Runs~3 and 4 (about 2.8 orbital
periods).
\footnote{One might alternatively estimate $P$ and $\Delta P$ (and
hence $\Delta E$) from the number of cycles contained in a given run
and the time resolution of the observations. However, with this method
we still predict $\Delta E \simeq 2.8$~cycles, even under the
optimistic assumption that an entire run could be used as the baseline
in the required cycle counts.}

We have also tried to establish more securely whether the 29-s
oscillations were present during the August (G160L) observing
sequences (Runs~1 and 2) despite the absence of any corresponding 
features in their power spectra in Figure~\ref{fig-lightpower}. To 
this end, we first high-pass filtered the first order white-light
light curves for these runs in a fashion identical to that used for the 
November (PRISM) time series. At phases away from eclipse, the
filtered first order light curves showed fluctuations of $\sim$1.5\%, 
which is consistent with the noise level in the unfiltered time 
series. We nevertheless proceeded 
to compute power spectra for the filtered G160L light curves, 
since sufficiently coherent periodic signals can sometimes be 
detected even if the signal-to-noise ratio is $<1$
\cite{scargle1}. However, no significant peaks corresponding to a 29~s
period were found in these power spectra. By injecting artificial signals
into the unfiltered time series, and analyzing the resulting fake data
in the same way as the actual observations, we determined that we
could have marginally (easily) detected a coherent sinusoidal
modulation with amplitude 0.5\% (1\%) at this frequency. 

As a final check, we also filtered and analyzed the zeroth order light
curves in the same fashion. A (weak) signal near the expected period
was seen only in the time series corresponding to the Run~2
post-eclipse data segment. Injecting artificial signals of
known strength into this data set, we found that the corresponding
oscillation amplitude was approximately 0.2\%. This is less than
the upper limit we are able to place on the oscillations in the UV
first order light curves, but comparable to the amplitude of the
oscillations around 3400~\AA~in the post-eclipse data segments of the
PRISM observations. (The latter statement is based on the oscillation
spectra presented in Section~\ref{oscspec}.) Given that this is the
{\em only} data segment in the G160L observations in which UX~UMa's
29-s oscillations may have been detected, we nevertheless conclude
that for most of the time covered by the August observing runs, the
oscillations must have been absent or weak compared to the November
sequences. This difference between the August and November
observations may be related to the fact that the system brightened 
by approximately 50\% between the two epochs (see
Section~\ref{discussion}. That DNO periods and amplitudes are quite
sensitive to changes in a system's brightness (and hence presumably
the mass transfer rate) is known from observations of DNe on the rise
or decline of an outburst (e.g. \citeNP{patterson1,hildebrand1}).

Finally, we note that some of the G160L power spectra did show a
consistent and (in at least one case) significant power excess at
somewhat shorter periods around 20~s. Judging from the size of 
the power peaks, the amplitude of the corresponding oscillations was 
probably just over 0.5\% but well below 1\%. Could these 20-s
oscillations in the August data be the counterpart of the 29-s
oscillations observed in November? This is certainly a tempting
identification, especially since WN72 and NR74 both established that
the period of the oscillations is not always the same. However, the
shortest period ever detected for the 29-s oscillations is 28.5 s
(NR74), significantly longer than the periods that are 
present in the G160L data sets. Selection effects could be 
responsible for this, and thus we cannot rule out that the two types of
oscillations do share a common origin. However, the current
observational database certainly does not yet establish any such
link. Because of their relative weakness, we will not consider these 
20-s oscillations further below.

\subsection{The Spectrum of the 29-s Oscillations}
\label{oscspec}

To isolate the spectra of the 29-s oscillations, we first subtracted a
5-point (about~1~oscillation cycle) running mean from each wavelength
pixel at each orbital phase separately for pre- ($\phi<0.95$), 
mid- ($0.98<\phi<1.02$) and post-eclipse ($\phi>1.05$) time
intervals. The resulting flux difference was then added to either 
a positive or negative oscillation spectrum, depending on the sign of
the flux datum at the same orbital phase in the similarly high-pass
filtered white light light curve. After dividing the co-added positive
and negative oscillation spectra by the number of orbital phase points
used in their construction, the negative mean oscillation spectra were
subtracted from the positive ones to yield the net oscillation
spectra. Since the mean pulse shape is highly sinusoidal, the
oscillation spectra were finally normalized to the peak-to-peak
oscillation amplitude by multiplying the monochromatic fluxes by a
factor of $\pi/2$ (the mean value of the positive/negative part of a
sinusoidal pulse with zero mean and unit amplitude is $\pm$2/$\pi$).

To characterize the oscillation spectra, we carried out $\chi^2$
fits to the data using three different types of models: (1) a single
temperature blackbody (BB); (2) a single temperature model
stellar atmosphere (SA); (3) the derivative of the BB spectrum 
with respect to temperature, $dB/dT$. The SA models were calculated 
with Hubeny's spectral synthesis code {\sc synspec} \cite{hubeny2} 
from {\sc atlas} \nocite{kurucz2} (for $T_{eff} \leq 50,000~K$) or
{\sc tlusty} \cite{hubeny1,hubeny3} (for $50,000~K < T_{eff} \leq
140,000~K$) structure models. The overall range of gravities on our
model grid was 2.0~$<$~log~g~$<$~7.0, though not all gravities were 
available at all temperatures. In the actual fits to the data, we 
found that models with the highest available gravity at a given 
temperature tended to be preferred (but note that model spectra 
tend to be more sensitive to temperature than to gravity in the 
relevant parameters regime). All model spectra were smoothed to
the instrumental resolution before comparing them to the data, and
E(B-V)=0.0 was assumed in all fits (c.f. Paper~I).

Models (1) and (2) are appropriate if the observed oscillations 
have their origin in a more or less constant temperature emitting
region whose projected area, as seen from Earth, is varying 
in a periodic fashion. Model (3) would be preferred if the
oscillations are due to a source presenting us with roughly 
constant projected area (away from eclipses, at least), 
but fluctuating in temperature. Note that while
the normalization constant required to match models (1) and (2) 
to the data is
simply proportional to the ratio of projected area and distance
squared, the same normalizing factor for model (3) includes an
additional multiplicative term which is equal to the magnitude of the
temperature fluctuations, $\Delta T$. Thus while models (1) and (2)
can be used straightforwardly to place constraints on the size of the 
emitting region, the same is not true of model (3).

The pre-, mid- and post-eclipse oscillation spectra constructed from
the Run~3 and 4 observing sequences are plotted along with the
best-fitting models in
Figures~\ref{fig-oscfits1}~and~\ref{fig-oscfits2}. 
The parameters/limits derived from
the model fits to the oscillation spectra are listed in
Table~\ref{tbl-fits}. The spectrum of the 
oscillations is extremely blue at orbital phases away from
eclipse. The model fits to these data all 
converge on effectively infinite temperatures, suggesting that we 
may be seeing the Rayleigh-Jeans tail of the oscillation
spectrum. [Note that the $dB/dT$ model also has Rayleigh-Jeans-like 
tail: in the limit $(h\nu)/(kT) \rightarrow 0$, $dB/dT \propto B(T)/T$ 
asymptotically, where all symbols have their usual meanings.] 
Our model fits nevertheless 
allow us to derive lower limits on the temperature of the source of
the oscillations at these phases. Inspection of Table~\ref{tbl-fits}
shows that $T >$~130,000~K, 100,000~K and 50,000~K at the $2\sigma$
level for all BB, SA and $dB/dT$ models, respectively. For the BB and SA
models, the lower limits on the temperature of the source can be
directly transformed into upper limits on its projected area,
$A_{proj}$. These turn out to be $A_{proj}/A_{WD} <$~0.01 and 0.02
for all BB and SA models, respectively, where $A_{WD}=4\pi R_{WD}^2$ 
is the surface area of the WD and a distance of 345~pc to UX~UMa has
been assumed \cite{baptista1}. We conclude that if our optically
thick, thermal models are appropriate, the oscillation light away from
eclipse is dominated by a very compact and extremely hot source.

The insets in the top panels of
Figures~\ref{fig-oscfits1}~and~\ref{fig-oscfits2} show the ratio of 
the pre-eclipse oscillation spectra to the total average pre-eclipse
spectra (c.f. Paper~I) as a function of
wavelength. These plots demonstrate explicitly that the colors of
the oscillations are much bluer than those of the disk and the
bright spot. Specifically, the peak-to-peak amplitude of the
oscillations at pre-eclipse orbital phases rises from well under 
1\% of the total light at long optical wavelengths to 
$\gtappeq 2$\% near 2000~\AA. This wavelength dependence may be the 
reason why the amplitude of the oscillations in our white-light 
PRISM light curves ($\simeq$ 0.5\%) is somewhat larger than in the 
ground-based optical observations of WN72 and NR74, since the latter
were not sensitive to the shortest wavelengths covered by the PRISM. 
As a simple test of this idea, we have estimated
the mean strength of the oscillations in the wavelength region
3000~\AA~-~8000~\AA~from the Run~3 pre-eclipse data segment (in which
the oscillations are strongest). The expected amplitude of the
oscillations in that bandpass is about 0.25\%, close to that found by
WN72 and NR74.

The mid-eclipse oscillation spectra in
Figures~\ref{fig-oscfits1}~and~\ref{fig-oscfits2} are 
rather noisy, but nevertheless markedly redder than the pre- and
post-eclipse oscillation spectra. Correspondingly, model fits to these
data yield much cooler temperatures in the range 
$20,000$~K$\ltappeq T \ltappeq 30,000$~K (BB and SA models) or
$10,000$~K$\ltappeq T \ltappeq 20,000$~K ($dB/dT$ model). The 
projected areas corresponding to the BB and SA model fits are of the 
order of a few (2-7) percent of the WD surface area. Note that all 
of our estimates are subject to revision if the pulsed emission 
is optically thin or is due to reflection of light from a hidden source 
(see below). In particular, the emitting region could then be much 
larger than suggested by the BB and SA model fits. In any 
event, what remains of the oscillation light at mid-eclipse appears to
be coming from a region that is cooler and more extended than the
source which dominates the oscillations at orbital phases away from
eclipse.

\section{Discussion -- The Origin of the 29-s Oscillations}
\label{discussion}

The properties of UX~UMa's 29-s oscillations place fairly
stringent constraints on models for their origin. This is important in
its own right, but assumes a more general significance because, as
noted in the Section~\ref{introduction}, DNOs are observed in a fair
number of DNe and a few other NLs (e.g. \citeNP{warner6}). 

Most fundamentally, it seems likely that the short time-scale of
the oscillations corresponds to the dynamical (rotational) time-scale
near the source of the oscillations. For UX~UMa's system parameters
($R_{WD} = 0.014R_{\sun}$, $M_{WD} = 0.47M_{\sun}$; 
\citeNP{baptista1}), 29~s corresponds to the Keplerian rotation period at
a radius of about 1.1~$R_{WD}$~in the accretion disk, i.e. very close
to the WD. The very blue spectrum of the oscillations away from
eclipse similarly suggests that their main source is very compact  
and extremely hot. These constraints are consistent with an origin of
the oscillations either in a small hot spot on a fast rotating WD, or
in a BL between the inner edge of the disk and the WD, but they would
seem to pose serious problems for any model in which the source of the
oscillations is identified with a disturbance further out in the
accretion disk. On the other hand, the shapes of the oscillation
eclipses -- specifically their non-totality and gradual 
in- and egresses -- demand that not all of the emission can come from
near the center of the disk. This is because these regions, including
the entire WD, are completely and rather abruptly occulted by the
secondary near mid-eclipse.

We are thus already forced to consider a two component model for the
origin of the observed modulations. The first of these components 
dominates the light at orbital phases away from eclipse and is
probably due to a compact emitting source near disk center. This
source is fully eclipsed by the secondary near conjunction. The 
second component is probably due to reprocessing of the light 
emitted by the compact source in the accretion disk atmosphere, 
and is not fully occulted even at mid-eclipse. The suppression of 
the oscillation amplitudes at post-eclipse orbital phases also 
seems to implicate 
the bright spot region at the disk edge as a possible reprocessing 
site. However, this interpretation is not unique, since 
some kind of post-eclipse absorption event could also produce the
observed asymmetries in the light curves of both the oscillations and
the total light. The same interpretive difficulty was noted in a
different context in Paper~I. 

Note that we use the term ``reprocessing'' loosely here, in the 
sense that the corresponding spectral component could in 
principle be due to either thermalization or reflection of 
the pulsed light emitted by the compact source. It might 
also represent recombination radiation emitted in response 
to photoionization of material in the disk atmosphere by 
the hot, compact source. 

A two component model such as that sketched above is also required to
account for the difference between the oscillations spectra away from
and during eclipse. The former are very blue and appear to be produced
by a compact, hot source. By contrast, the latter are much redder and
seem to arise in a cooler and more extended region. Note, however,
that the area estimates provided by our BB and SA model fits to the 
mid-eclipse oscillation spectra, while larger than the corresponding
estimates for the out-of-eclipse spectra, are still only a few percent
of the WD surface area. This is much smaller than the portion of
the accretion disk that is visible at mid-eclipse. (Based on UX~UMa's
system parameters, we estimate that more than half of the disk surface
area remains unocculted at all times.) It should be kept in mind,
however, that the models we have used to fit and characterize the data
are optically thick, whereas at least the reprocessing component revealed 
in the mid-eclipse oscillation spectra might not be. If this component
is in fact optically thin, the reprocessing site could be much larger 
than suggested by our fits. We can also not exclude the possibility 
that the hot, ``compact'' component we observe could itself be due to 
reflection. In this case the ``true'' source of the oscillations 
must be hidden from view and could also be larger than indicated 
by our model fits to the out-of-eclipse spectra.

A related potential problem faced by this type of model is the
apparent dichotomy between the very blue out-of-eclipse oscillation 
spectrum -- which might be taken to indicate that the compact, hot
component dominates almost completely at these orbital phases
-- and the rather gradual eclipses of the oscillations in
Figure~\ref{fig-thirty} -- which suggest that a relatively large 
area contributes significantly to the oscillations 
away from eclipse. However, we have not yet tried to fit any two
component models to the out-of-eclipse oscillation spectra. (This is
partly because the $\chi^2$ values produced by our single component
model fits to these relatively noisy data were already low, and
partly because such an analysis lies beyond the scope of the present 
paper.) Consequently, we are unable to set a limit on the possible
contribution of the extended, reprocessing 
component to the oscillating flux away from eclipse. In the absence of
such a constraint, we cannot tell if there might be a conflict
between the data and a simple two component model. 

A qualitative constraint on the relative strength of the two
components may be derived from the fact that the oscillations lose one
cycle during eclipse. Quite generally, if the 29-s oscillations are
associated with prograde rotation of the emitting source(s), a
significant and probably dominant fraction of the emission must be
beamed in such a way as to produce pulse maxima when the corresponding
emitter is nearest to us. Only then will successive pulse maxima occur
later and later after the limb of the secondary first passes over 
the line of centers of the two stars in the system. An obvious example
of an emitting source meeting this constraint would be a bright 
spot on the surface of a fast rotating WD \cite{petterson1}. By contrast, 
if the oscillations were dominated by the reprocessing of
light emitted by such a spot in the atmosphere of a concave, 
optically thick accretion
disk, pulse maxima would be expected to occur when the spot is 
illuminating the far side of the disk, which is less obliquely 
inclined towards our line of sight. In that case, the oscillations
would gain a cycle during every eclipse, a
situation that is actually encountered in the IP 
DQ~Her (e.g. \citeNP{zhang2}). Thus with a two component model 
including a directly observed and a reprocessed component, both 
positive and
negative eclipse phase shifts can be accounted for. In the general
scenario, in which both the compact source of direct light and the
extended reprocessing disk contribute significantly to the
modulated flux, the sign of the observed phase shift will depend on
the relative strength of these two components \cite{petterson1}. 
The negative phase shift in UX~UMa thus suggests that the
direct, compact component contributes the majority of the
oscillation light away from eclipse, provided that the optically 
thick accretion disk is the main reprocessing site.

A final constraint on the origin of the oscillations can be 
derived from the observed pulse shape away from eclipse
(Figure~\ref{fig-sinefit}). Let us assume for the moment that the
oscillations at these orbital phases are dominated 
by direct light from a rotating, compact source which 
presents us with varying amounts of projected area over the course of
an oscillation cycle. More specifically, we will identify this
compact source with either a bright spot on the surface of a  
fast rotating WD or with a localized disturbance in an equatorial BL
at the inner disk edge. Now the highly sinusoidal pulse shape implies that
this compact source is never fully occulted during the oscillation 
cycle. This rules out any site for the compact source lying within
$\ltappeq 3~R_{WD}$ of disk center in the equatorial plane, since all
locations closer in will be occulted once during every cycle by the body of the
WD. Thus the identification of the compact source with a disturbance
in the inner disk or an equatorial BL is problematic. Similarly, 
Petterson's~(1980) numerical model for the 29-s oscillations -- which
is essentially identical to our two component (direct + reprocessed)
model, but explicitly relies on bright spots on the equator of the
rotating WD -- is ruled out by the new data since it predicts a
non-sinusoidal pulse shape.

Can magnetically controlled accretion near the WD provide a more
promising alternative? In the simplest version of this picture, the
magnetic field of the WD is strong enough to disrupt the inner disk
and force the accreting material to flow along field lines onto one or
both magnetic polecaps. The impact of the accretion flow onto the
poles produces one or two bright spots on the WD surface which
can then be identified with the compact source that dominates the
observed 29-s oscillations away from eclipse. This model is
essentially just a weak field IP scenario for UX~UMa.

If both accreting poles were visible to us, the period of the
oscillations would correspond to one half of the spin period of the WD
or, equivalently, to one half of the rotation period of material near
the inner disk edge. However, this would imply an inner disk radius of
1.8~$R_{WD}$, whereas a hole extending out to $3.1~R_{WD}$
would be required for the second pole to be unocculted by the
optically thick inner disk. Thus only one of the magnetic poles can
actually be visible in UX~UMa. It is nevertheless easy to explain the
sinusoidal pulse shape within an IP framework, since it is only
required that the inclination of the magnetic axis with respect to the
disk and WD rotation axes be small enough for the visible accreting
pole to avoid self-eclipses. For UX~UMa's orbital inclination of
i=71$^o$ \cite{baptista1}, the maximum allowed inclination of the 
magnetic axis is 90$^o$-i=19$^o$.

In an IP model with a single visible pole, we can identify the
oscillation period directly with the WD spin period and the Keplerian
period near the inner disk edge. Thus the accretion disk must extend
down to about $1.1~R_{WD}$ before magnetic stresses disrupt it.
\footnote{Note that if the the period of the oscillations corresponded
to the spin period of the WD, but the inner edge of the disk were
located much farther out than 1.1~$R_{WD}$, material at the inner disk
edge would be rotating more slowly than the field lines it is trying to
latch onto. It would therefore be repelled out to larger radii and
perhaps even out of the system by this centrifugal barrier. This
propeller mechanism is thought to operate in the intermediate polar
AE~Aqr \cite{erac1}.}
This may be identified with the radius of
the WD magnetosphere (the Alfv\'{e}n radius), which for a disk
threaded by a dipolar field is located at roughly \cite{frank1}
\begin{equation}
R_{A}\simeq 2.5 \times 10^8 \dot{M}_{16}^{-2/7} M_{WD}^{-1/7}
\mu_{30}^{4/7} cm
\end{equation}
where the parameters have been scaled to units appropriate for CVs,
i.e. $M_{WD}$ is the mass of the accreting WD in solar masses,
$\dot{M}_{16}$ is the accretion rate in units of $10^{16}$~g~s$^{-1}$
and $\mu_{30}$ is the magnetic moment of the WD in units of
$10^{30}$~G~cm$^3$. Taking UX~UMa's system parameters, $\dot{M}_{16}
\simeq 10-100$, and setting $R_A=1.1R_{WD}$, we find $\mu_{30} \simeq
30-100$, at the low end of the range for IPs, as it must be
\cite{warner6}.

However, there are two objections to the direct application of an IP
model to UX~UMa. The first is that it requires the WD to be rotating
at more than 90\% of break-up, which might be considered to be a 
uncomfortably high value. On its own, this objection could actually be
turned around to provide an explanation for the
``missing BL'' problem in ``non-magnetic'' CVs (see
\citeNP{hoare1,hoare2} and references therein). \footnote{In a
magnetic accretion scenario there would of course be no classical BL
anyway, but unless the WD is a rapid rotator, roughly one half of the
accretion luminosity would still have to be released from a very small
region at the center of the accretion disk. Thus the ``missing BL''
problem would simply become a ``missing flux'' problem in this case,
which is probably a more appropriate view in any case.}

The second objection is more potent and arises from the changes in the
oscillation period that were observed by NR74. For example, during one
of their multi-night observing sequences, the period changed roughly
monotonically from 30.0~s to 28.5~s within the course of only 5
days. This corresponds to a period derivative $\dot{P} \sim -3\times
10^{-6}$~s~s$^{-1}$ and a time-scale for period changes of
$\tau_{spin-up} = P/(-\dot{P}) \simeq 8 \times 10^{6}$~s~$\sim
100$~days. This should be compared to the minimum spin-up time-scale
that can be achieved by any type of accretion (magnetic or
otherwise). This minimum time-scale is obtained by considering the
time required to spin-up the WD from rest to break-up if {\em all} of
the angular momentum of the accreting matter goes into WD rotation. A
simple calculation gives
\begin{equation}
\tau_{spin-up} \sim \frac{M_{WD}}{\dot{M}_{acc}}
\left(\frac{R_{WD}}{R_{circ}}\right)^{1/2}
\label{eq-spinup}
\end{equation}
where $R_{circ}$ is the circularization radius of of the accreting
material. For UX~UMa's system parameters, $R_{circ} \simeq 12.5R_{WD}$
and $\tau_{spin-up} \sim 10^{7}$~years for reasonable accretion rates
between $10^{17}-10^{18}$~g~s$^{-1}$. This tremendous discrepancy with
the time-scale over which the oscillation period is seen to vary shows
that the observed period changes in UX~UMa cannot possibly reflect
actual WD spin period variations in response to accretion. Thus the
simple IP model for UX~UMa fails.

Warner~(1995) has recently proposed a qualitative unified scheme for
DNOs, which may offer a way out of this dilemma. Building on a model
described originally by \citeN{paczynski2} (see also \citeNP{king3}), 
he suggested that DNOs occur as a result of magnetic accretion in
systems in which the WD magnetic field is weak enough to permit the
outer layers of the WD to rotate differentially. In this case, the 
observed DNO periods correspond not to the spin period of the 
entire WD, but to the rotation period of its surface layers. The 
observed spin-up time scales can then be explained, since 
$M_{WD}$ in Equation~\ref{eq-spinup} can now be replaced by the 
mass of the rotating layers only (which must be $\ltappeq
10^{-8}M_{\sun}$). Thus, changes in the oscillation periods can
plausibly be accounted for by accretion rate variations in this 
model. Since magnetic accretion will only occur at all if
$R_{A}>R_{WD}$, DNOs should not be present in all ``non-magnetic''
CVs, as is observed. Moreover, relatively more DNe than NLs should
exhibit DNOs, since the latter will on average have higher mean
accretion rates and smaller magnetospheres. This, too, is in line 
with the available data.

What about the apparent absence of the oscillations during the August
observations, when the system was 50\%~fainter than in November? 
A qualitative explanation for this behaviour is suggested by outburst
observations of the dwarf nova AH~Her \cite{hildebrand1}. In this system, DNOs
are present on the rising and declining outburst branches, with
amplitudes (periods) that increase (decrease) with increasing
brightness. However, near maximum light at the peak of the outburst,
the DNOs suddenly disappear. In the context of a magnetic accretion
model, this suggests that the magnetosphere shrinks in response to an
increase in the accretion rate just as expected. The growth of the DNO
amplitudes occurs because more accretion energy becomes available for
generating the oscillations as the disk-magnetosphere interaction
region moves towards smaller radii. The DNO periods decrease because
the Keplerian velocities increase towards disk center. These trends 
continue until the magnetosphere is crushed onto the WD surface, at
which point the DNOs cease abruptly. Thus an increase in DNO amplitude
with increasing accretion rate, as suggested by UX~UMa, need not be 
in conflict with a magnetic accretion model. 

Despite this apparent success, we feel it is too early to accept a
model of this type for the origin of DNOs. Most importantly, we are
not sure what magnetic accretion onto differentially rotating WD
surface layers really means physically. For example, it is not clear
{\em a priori} whether the magnetic field controlling the accretion
flow should be thought of as 
stable and anchored in the slowly rotating WD core or, alternatively,
as transient and perhaps generated, as well
as anchored, in the differentially rotating surface layers themselves
(possibly by some sort of dynamo action; e.g. \citeNP{king3}). Partly
as a result of this uncertainty, it is also not obvious what type of 
accretion geometry should be expected in this picture close to 
the WD surface. Thus more theoretical work will be required 
before a magnetic accretion scenario of this kind may be judged
successful.

As an incentive for theoreticians to tackle this problem, we conclude
this section by placing UX~UMa's 29-s oscillations in the context of
DNOs more generally. Combining data from Warner~(1995) and
\citeN{ritter1}, we plot in the bottom left panel of
Figure~\ref{fig-alldno} DNO period against WD mass for all systems in
which DNOs are observed and for which WD masses are available. Each
continuous curve in Figure~\ref{fig-alldno} gives the Keplerian
rotation periods in an accretion disks around a primary of the given
mass at the indicated radius in the disk. The analytical approximation
of \citeN{nauenberg1} to the \citeN{hamada1} mass-radius
relationship for cool degenerate stars has been assumed in deriving 
these curves. Where vertical error bars are shown, they correspond to
the range of periods that have been observed in the system. 

Several points are worth noting from this plot. First,
the periods of all DNOs correspond to the Keplerian time-scales at
radii $R_{DNO} \simeq 1-3~R_{WD}$ in the accretion disk. In no case
are the DNO periods inconsistent with the basic requirement that
$R_{DNO} > R_{WD}$. Second, several DNe exhibit period variations
similar to or larger than those noted above for UX~UMa. Thus the
identification of DNO periods with WD spin periods is untenable in 
these systems also. Third, the two IPs that might be said to exhibit
DNOs (AE~Aqr with $P_{DNO} = 33$~s and DQ~Her with $P_{DNO} = 71$~s)
both occupy the region close to the $R_{DNO} = 3~R_{WD}$ line in
Figure~\ref{fig-alldno}. This would appear to be consistent with a
magnetic accretion scenario since the field strength (and hence
magnetospheres) should be larger in these systems than in DNe and
NLs. However, in both of these systems the oscillation periods are
extremely stable and almost certainly do correspond to the true WD
spin period (or perhaps one half of it, in the case of DQ~Her;
\citeNP{zhang2}). Fourth and finally, there may be a hint of
clustering near $P_{DNO} \simeq 30$~s in Figure~\ref{fig-alldno}.

This last property is seen more easily in a simple histogram of the
observational DNO-period distribution function which we show in the
bottom right panel of Figure~\ref{fig-alldno}. The cluster of CVs with
$P_{DNO} \simeq 30$~s 
is fairly obvious in this figure. For comparison, the WD mass
distribution function for these systems is plotted in the top left panel
of Figure~\ref{fig-alldno}. This histogram does not seem to show 
a similarly sharp peak, but the relatively large errors on the WD
mass estimates may be partly responsible for this. (A clustering of 
oscillation periods could be easily explained
by any model in which $P_{DNO}$ is set by the dynamical time-scale
near the WD {\em if} the WD mass distribution function showed a similar
peak.) In any case, the currently known number of CVs that exhibit
DNOs and have sufficiently well established periods is probably too
small for the peak in the DNO-period function to attain 
statistical significance. However, if the clustering of oscillation
periods around 30~s is confirmed in the future and shown {\em not} to
arise simply from the underlying WD mass distribution function, this
property will have to be accounted for by any successful model for
the origin of DNOs.

\section{Conclusions}
\label{conclusions}

Low amplitude ($\simeq 0.5\%$) coherent 29-s oscillations have been
detected in HST/FOS eclipse observations of the NL variable
UX~UMa. These are the same DN-type oscillations that were originally
discovered in this system by WN72 and subsequently analyzed in more
detail by NR74. The oscillations are easily seen in one pair of
eclipse sequences obtained with the FOS/PRISM in November of 
1994, but not in a similar pair obtained with the FOS/G160L
grating in August of the same year (except, perhaps, in one isolated
data segment of the zeroth order light curves; see
Section~\ref{waveave}).

We find that the oscillations in the PRISM data are sinusoidal to
within the small observational errors and undergo an approximately
$-360^o$ phase shift during eclipses (i.e. one cycle is lost). These
results are similar to those derived by WN72 and NR74. We also detect
orbital phase-related amplitude variations in the oscillation time
series. Specifically, the oscillation amplitudes are highest at
pre-eclipse orbital phases and exhibit a rather gradual eclipse 
whose shape is roughly similar to, though perhaps slightly narrower
than UX~UMa's overall light curve in the PRISM
bandpass (2000~\AA~-~8000~\AA).

PRISM spectra of the oscillations constructed from data segments
covering pre- and post-eclipse orbital phases are 
extremely blue. Single component, optically thick model fits to these
data only allow limits to be placed on the source temperature
and size. In fits to the data with BB (SA) models, the lower limits on
temperature always turn out to be $\geq$~95,000~K ($\geq$~85,000~K),
and the corresponding upper limits on the projected area of 
the source are all $\leq 2\%$ of the WD surface area. 
Fits to the same data with the derivative of the blackbody spectrum, 
$dB/dT$ -- which would be appropriate if the observed oscillations
were due to a source fluctuating in temperature (rather than to a
source fluctuating in projected area) -- 
also tend to converge towards infinite temperatures and yield lower
limits $\geq$~50,000~K. Thus the spectra and model fits all suggest
that the source dominating the oscillations away from eclipse is
extremely hot and probably very compact.

By contrast, the two oscillation spectra derived from data segments
covering mid-eclipse are much redder. Correspondingly, model fits to
these data yield cooler temperature estimates in the range 
$20,000$~K$\ltappeq T \ltappeq 30,000$~K (BB and SA models) and 
$10,000$~K$\ltappeq T \ltappeq 20,000$~K ($dB/dT$ model). The projected 
areas corresponding to the BB and SA model fits to the mid-eclipse
oscillation spectra are of the order of a few percent of 
the WD surface area. Thus what remains of the oscillation light at
mid-eclipse appears to be coming from a region that is cooler and more
extended than the source which dominates the spectrum of the
oscillations at orbital phases away from eclipse. 

Based on these observational constraints, we suggest that the ultimate
source of the oscillations is probably a hot, compact region near disk
center, although significant reprocessing of the light emitted by
this source in the accretion disk and probably the BS must also take
place. This kind of two component (direct + reprocessed) model appears
to be able to account for all of the observed behavior, although it
remains to be seen whether the very blue oscillation spectra away from
eclipse can be reconciled quantitatively with the relatively broad and
gradual eclipses of the oscillations in this picture. It should also
be noted in this context that if the reprocessed emission is optically
thin, the reprocessing site may be much larger than suggested by our
optically thick model fits to the mid-eclipse oscillation spectra.

One {\em a priori} possible identification of the hot, compact 
source in this model is with a disturbance in the inner disk ar a
classical, equatorial BL. However, the highly sinusoidal pulse shape
of the oscillations does not permit this. The compact
source might instead be identified with a bright spot on the surface
of the rotating WD that, in an IP-type model for the origin of the
oscillations, may be associated with an accreting magnetic
pole. However, a standard
weak-field IP model for UX~UMa can also be ruled out, since WN72 and
NR74 observed the oscillation period to change on time-scales much
shorter than the minimum time-scale required to spin-up the WD by
accretion torques. A scenario along the lines recently 
proposed by Warner~(1995), in which the oscillations
arise as a result of magnetic accretion onto differentially rotating
WD surface layers, is still viable, but requires more theoretical 
work before it may be judged successful. 

We finally note that the characteristics of UX~UMa's 
oscillations place them quite squarely among DNOs in other CVs. In 
all systems, the period of the oscillations corresponds to the
dynamical time-scale in the accretion disk at $1-3R_{WD}$. 
There is a hint that DNO periods may cluster around $P_{DNO}
\simeq 30$~s despite the absence of a corresponding peak in the WD
mass distribution function. However, since the number of CVs with
established DNO periods and known WD masses is quite small, whereas
errors on WD masses are relatively large, the reality and
significance of this clustering needs to be confirmed.

\acknowledgements We gratefully acknowledge the support of NASA
through HST grant GO-5448 without which this work would not have been
possible. In addition, RAW acknowledges financial support from NASA
through grants NAGW-3171 and from STScI through grant GO-3683.03-91A,
both to the Pennsylvania State University. We would also like to thank
Chris Mauche, Ivan Hubeny and Rene Rutten for their contributions to 
this project, and the anonymous referee for a very helpful report.

\bibliographystyle{apj} \bibliography{bibliography}


\begin{deluxetable}{cccllc}
\tablewidth{500pt} 
\footnotesize 
\tablecaption{Results of model fits to the oscillation spectra.\label{tbl-fits}}
\tablehead{
\colhead{Spectrum} & 
\colhead{Run} & 
\colhead{Model}&
\colhead{~~T~~}& 
\colhead{$A_{proj}$\tablenotemark{a}}&
\colhead{$\chi^2_{\nu}$\tablenotemark{b}} \nl 
\colhead{} & 
\colhead{} &
\colhead{} & 
\colhead{($10^3$~K)}& 
\colhead{($A_{WD}$)}&
\colhead{(N=263)}\nl}
\startdata 
Pre-Eclipse & 3 & BB    & $>$230& $<$0.004 & 1.10 \nl
($\phi_{orb}<0.95$)
            & 3 & SA    & $>$105        & $<$0.02  & 1.13 \nl
            & 3 & $dB/dT$       & $>$80 &        & 1.10 \nl
            & 4 & BB    & $>$95 & $<$0.01 & 0.95 \nl
            & 4 & SA    & $>$85 & $<$0.02  & 0.97 \nl
            & 4 & $dB/dT$       & $>$50 &        & 0.95 \nl
Mid-Eclipse & 3 & BB    & 18$^{+20}_{-5}$& 0.05$^{+0.12}_{-0.04}$& 0.44 \nl
($0.98<\phi_{orb}<1.02$)
            & 3 & SA    & 19$^{+10}_{-4}$ & 0.07$^{+0.04}_{-0.05}$ & 0.44 \nl
            & 3 & $dB/dT$       & 13$^{+6}_{-3}$ &                        & 0.44 \nl
            & 4 & BB    & 28$^{+145}_{-11}$& 0.02$^{+0.05}_{-0.019}$& 0.41 \nl
            & 4 & SA    & 26$^{+114}_{-8}$& 0.03$^{+0.04}_{-0.028}$& 0.41 \nl
            & 4 & $dB/dT$       & 18$^{+17}_{-5}$&                        & 0.41 \nl
Post-Eclipse& 3 & BB    & $>$110        & $<$0.005 & 0.95 \nl
($\phi_{orb}<1.05$)
            & 3 & SA    & $>$85 & $<$0.01  & 0.97 \nl
            & 3 & $dB/dT$       & $>$60 &        & 0.95 \nl
            & 4 & BB    & $>$145        & $<$0.003 & 1.29 \nl
            & 4 & SA    & $>$95 & $<$0.009 & 1.31 \nl
            & 4 & $dB/dT$       & $>$65 &        & 1.29 \nl \hline
\enddata
\tablecomments{Errors and lower limits are 2$\sigma$ (95\% confidence).}
\tablenotetext{a}{$A_{proj}$ is the projected area implied by the fit
(in units of the WD surface area, $A_{WD}=4 \pi R_{WD}^{2}$) for an
assumed distance of d=345~pc and $R_{WD}=0.014 R_{\sun}$\cite{baptista1}.}
\tablenotetext{b}{Where only lower limits on temperatures are
given, the $\chi^{2}_{\nu}$ values correspond to fits with 
$T_{eff} > 10^9$~K (BB and $dB/dT$ models) or $T_{eff} > 1.4 \times 
10^5$~K (SA models).}
\normalsize
\end{deluxetable}

\clearpage

\figcaption[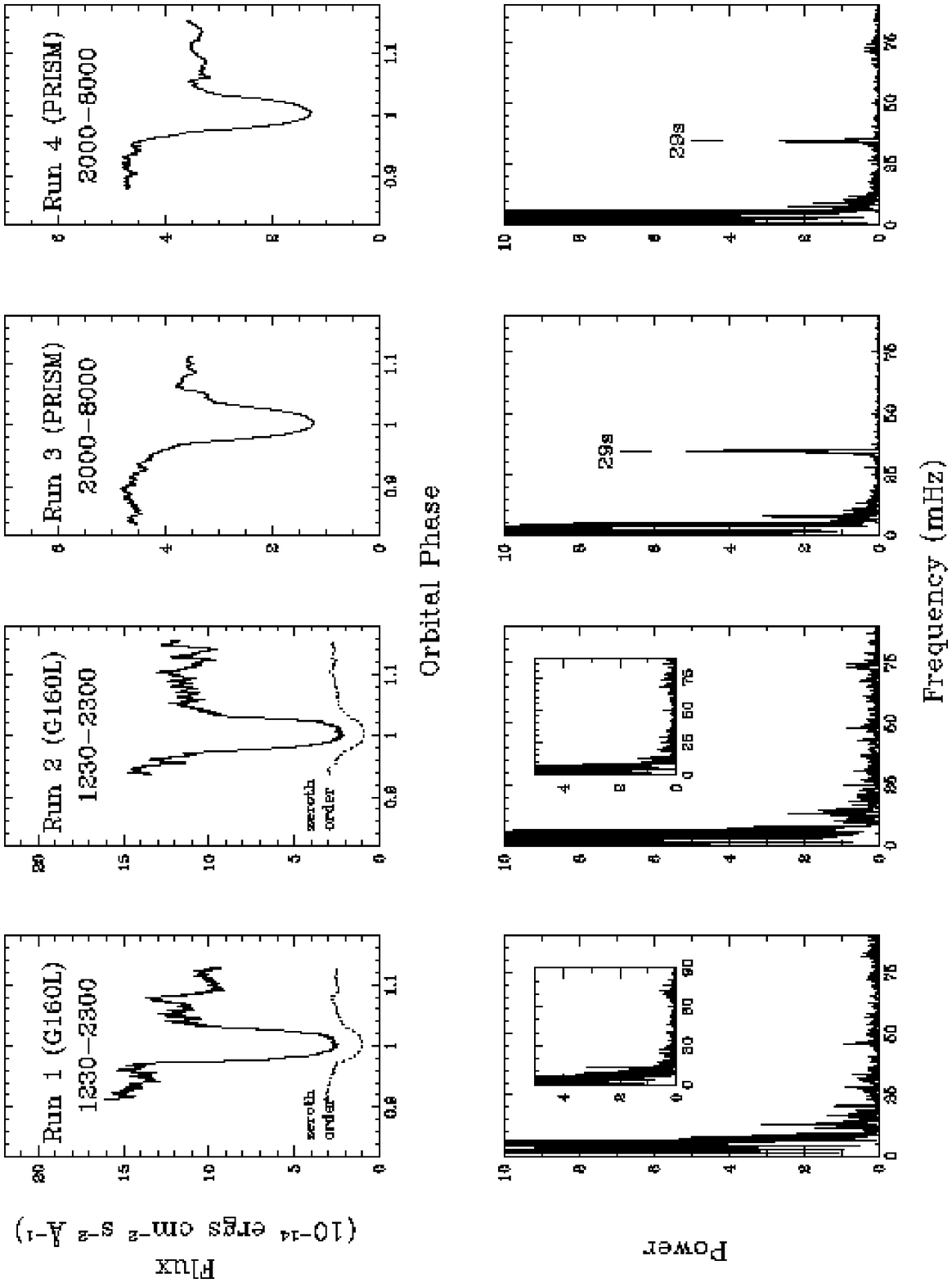]{Light curves and power spectra constructed
from the time-resolved HST/FOS spectra of UX~UMa. {\em Top panels:}
``White-light'' (solid lines) and zeroth order (dashed lines; Runs~1
and 2 only) light curves phased according to the ephemeris of
Baptista~{et al.} (1995). The observing sequence corresponding to
the light curve(s) in a given panel and the wavelength range over
which the spectra have been averaged to produce the first-order
light curve are indicated in each panel. See text for the calibration
of the zeroth order G160L light curves. {\em Bottom panels:}
Lomb-Scargle power spectra calculated from the the light curves shown
in the corresponding top panels; to minimize the effects of the
eclipses, light curves were divided by smoothed versions of themselves
before calculating the power spectra.\label{fig-lightpower}}

\figcaption[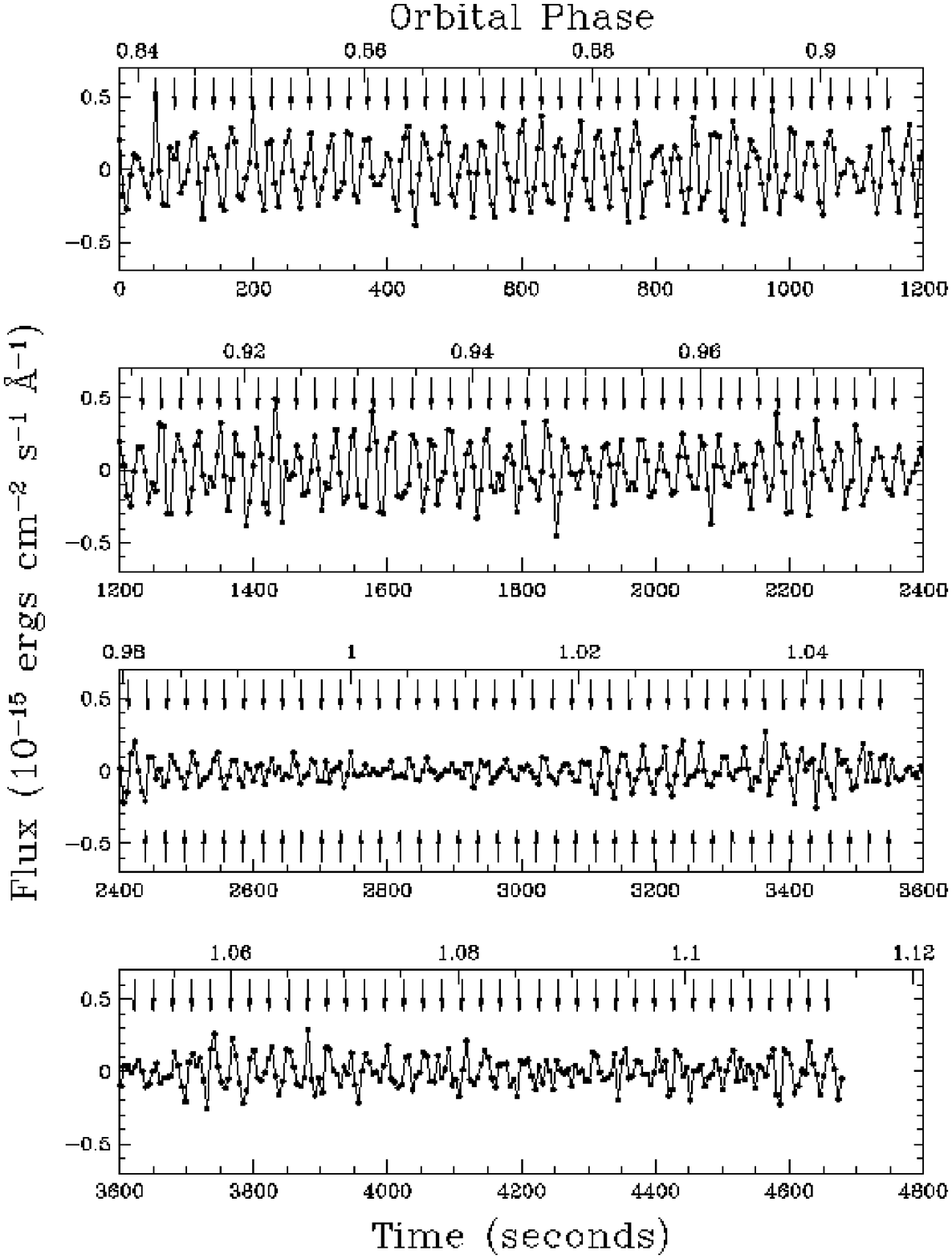]{The high-pass filtered time series for
Run~3. Shown is the white light (2000~\AA~-~8000~\AA)
light curve for Run~3 after subtraction of a 5-point running mean,
to reveal the small amplitude (approximately 0.5\%) 29-s
oscillations. The data 
are shown as a function of both time (in seconds; bottom x axes) and
orbital phase (top x axes). The arrows above the data mark the
times of maxima predicted from the peak in the pre-eclipse power 
spectrum after aligning to the first obvious observed maximum. The 
arrows below the data at phases close to conjunction
(third panel from the top) mark the times of minima predicted from 
the mid-eclipse power spectrum after aligning to the first obvious
observed minimum.\label{fig-cohere3}} 

\figcaption[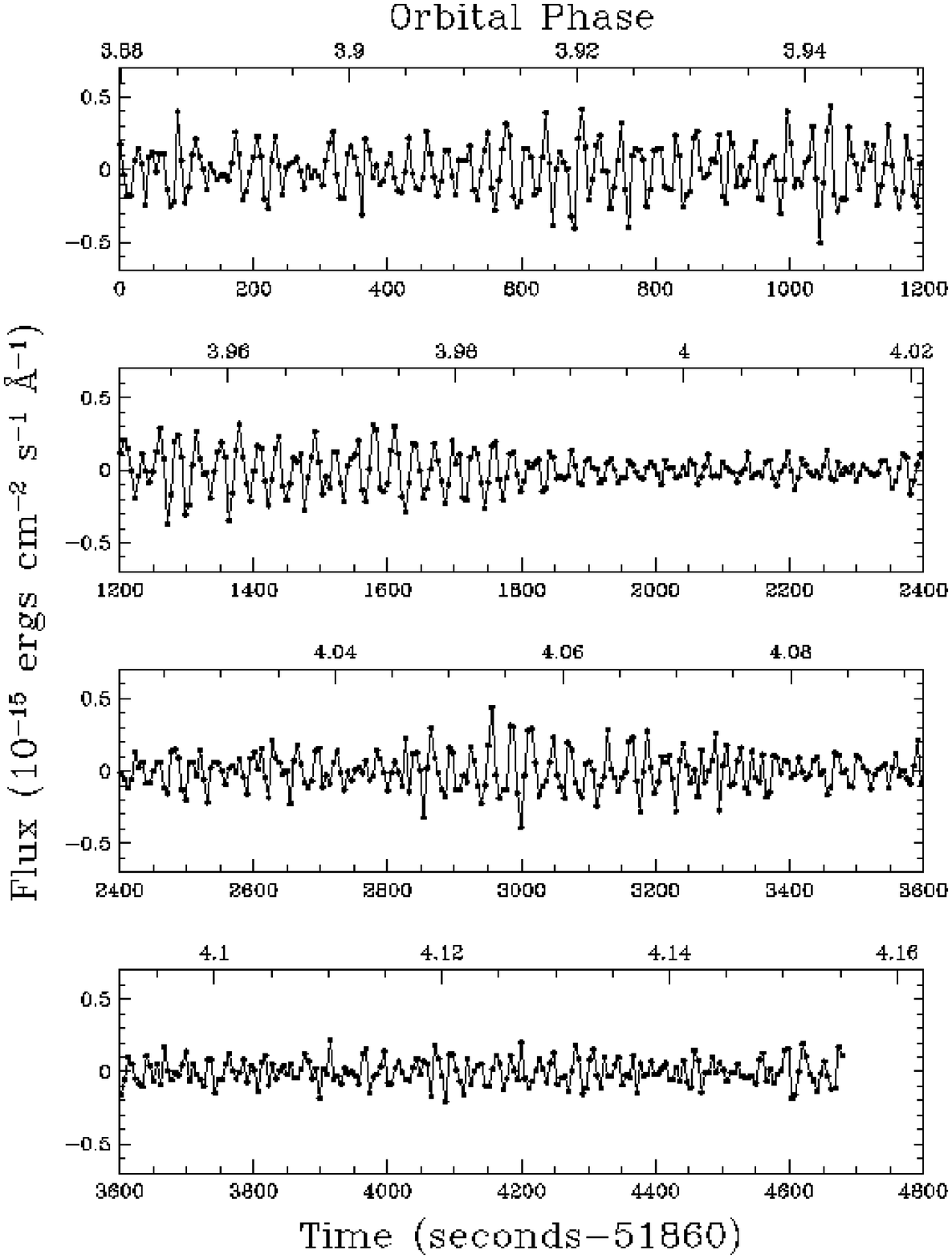]{As Figure~\ref{fig-cohere4}, but 
for the Run~4 data set.\label{fig-cohere4}} 

\figcaption[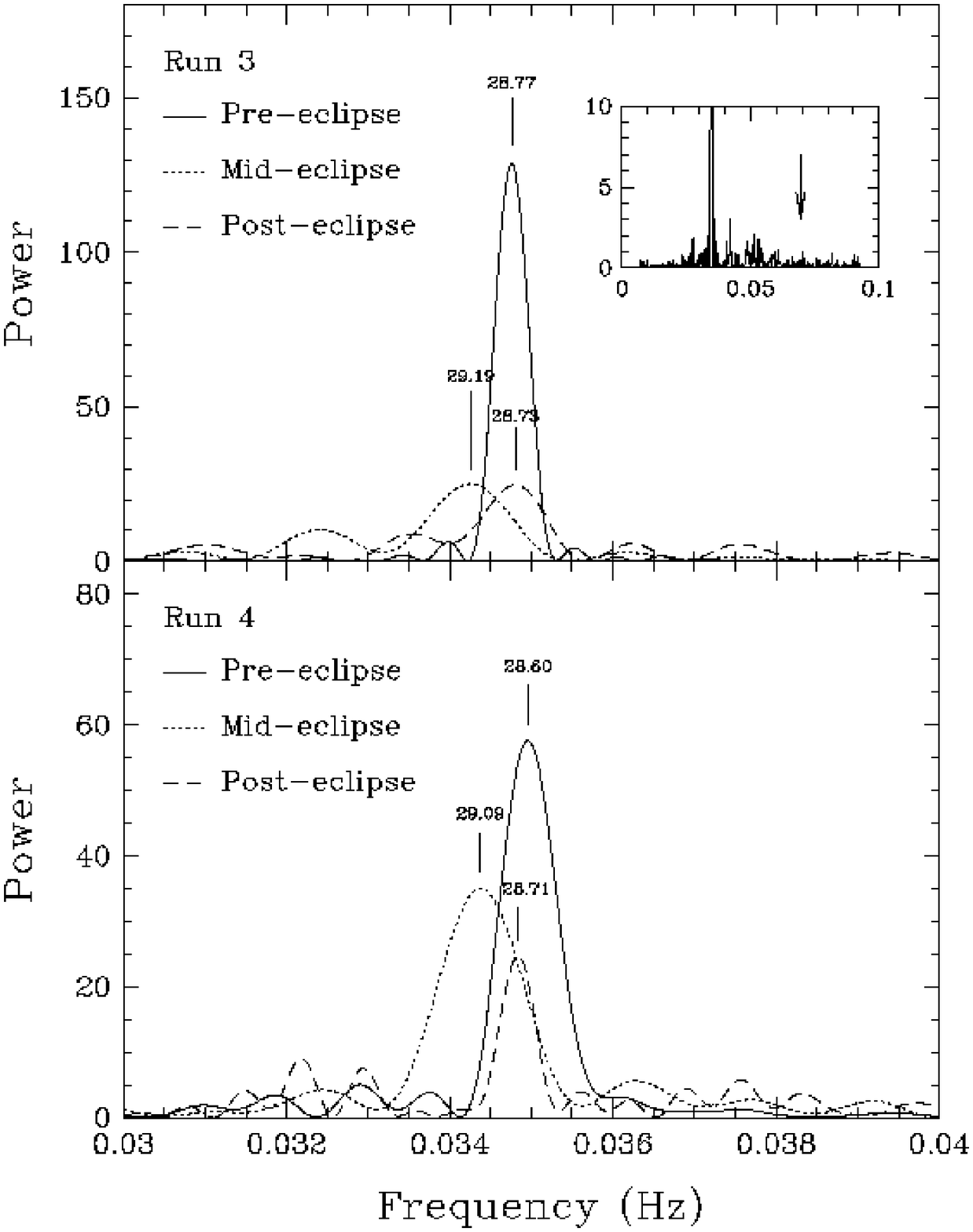]{Lomb-Scargle power spectra for the pre-,
mid- and post-eclipse segments of the high-pass filtered light curves
of Run~3 (top panel) and Run~4
(bottom panel). The peak in each power spectrum is marked with the
period (in seconds) to which it corresponds. The inset in the top
panel shows the pre-eclipse power spectrum for Run~3 over a larger
frequency and smaller power range. The arrow marks the position of 
the (absent) first overtone of the 28.77 s period in this part of the
data. \label{fig-power_30s}} 
  
\figcaption[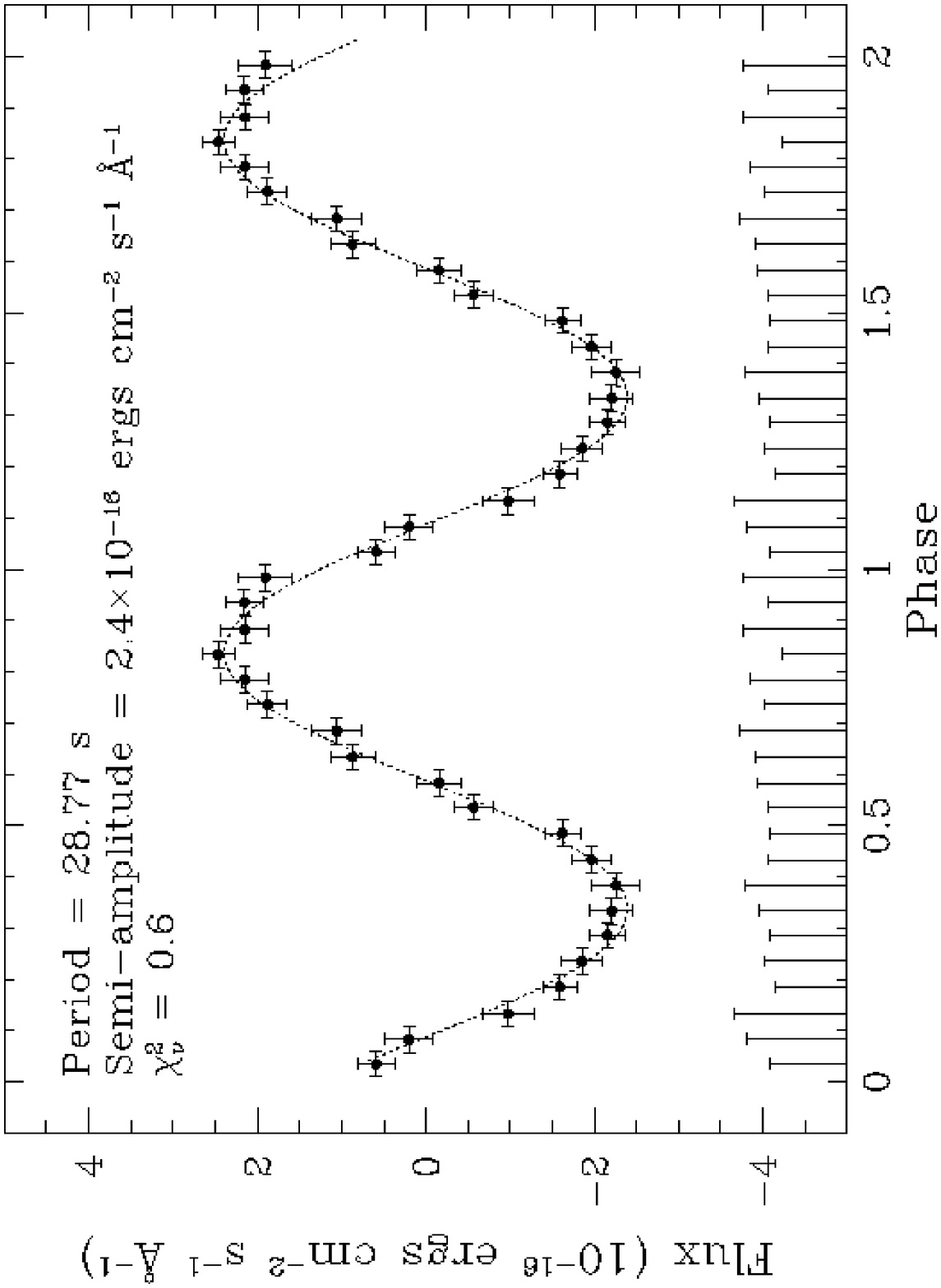]{The result of folding the pre-eclipse,
high-pass filtered white light time series derived from Run~3 onto 
the 28.77 s oscillation period. The data have been binned to a phase
resolution of 0.05 (marked by the horizontal error bars) and are shown
repeated over two oscillation cycles for clarity. The vertical
error bars correspond to the error on the mean of the data points in
each bin; the size of the vertical bars shown at the bottom axis
corresponds to the standard deviation of the points in each bin 
around the mean value. The dotted line is the best-fitting
fixed-period sinusoid for this data set. Based on its reduced 
$\chi^2$ of 0.6, the fit describes the data
acceptably.\label{fig-sinefit}}

\figcaption[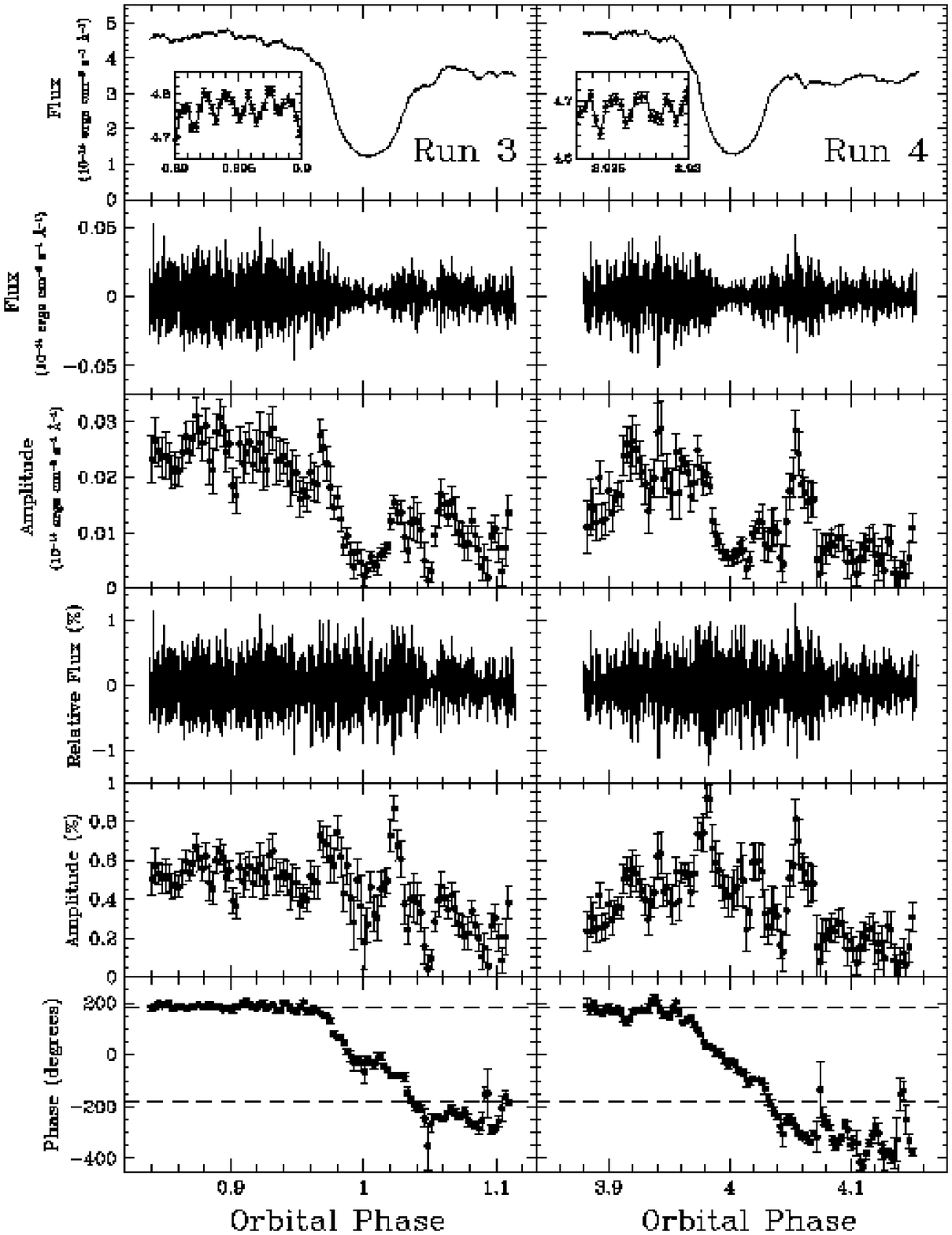]{The results of high-pass filtering and
demodulating the white light (2000~\AA~-~8000~\AA) light curves for
Runs~3 (left panels) and 4 (right panels). From top to bottom, we show
as a function of orbital phase: the unfiltered light curves (insets
zoom in on small portions of the data), the
high-pass filtered light curves, the oscillation amplitudes derived
from the latter, the high-pass filtered light curves expressed as
a percentage of the total light at a given phase, the oscillation
amplitudes derived from the latter and the phase of the
oscillations. The oscillation phases for both runs have been shifted 
to a mean value of 180$^o$ at pre-eclipse orbital phases (see
text). Only every other data points in the amplitude and phase plots
is independent.\label{fig-thirty}}

\figcaption[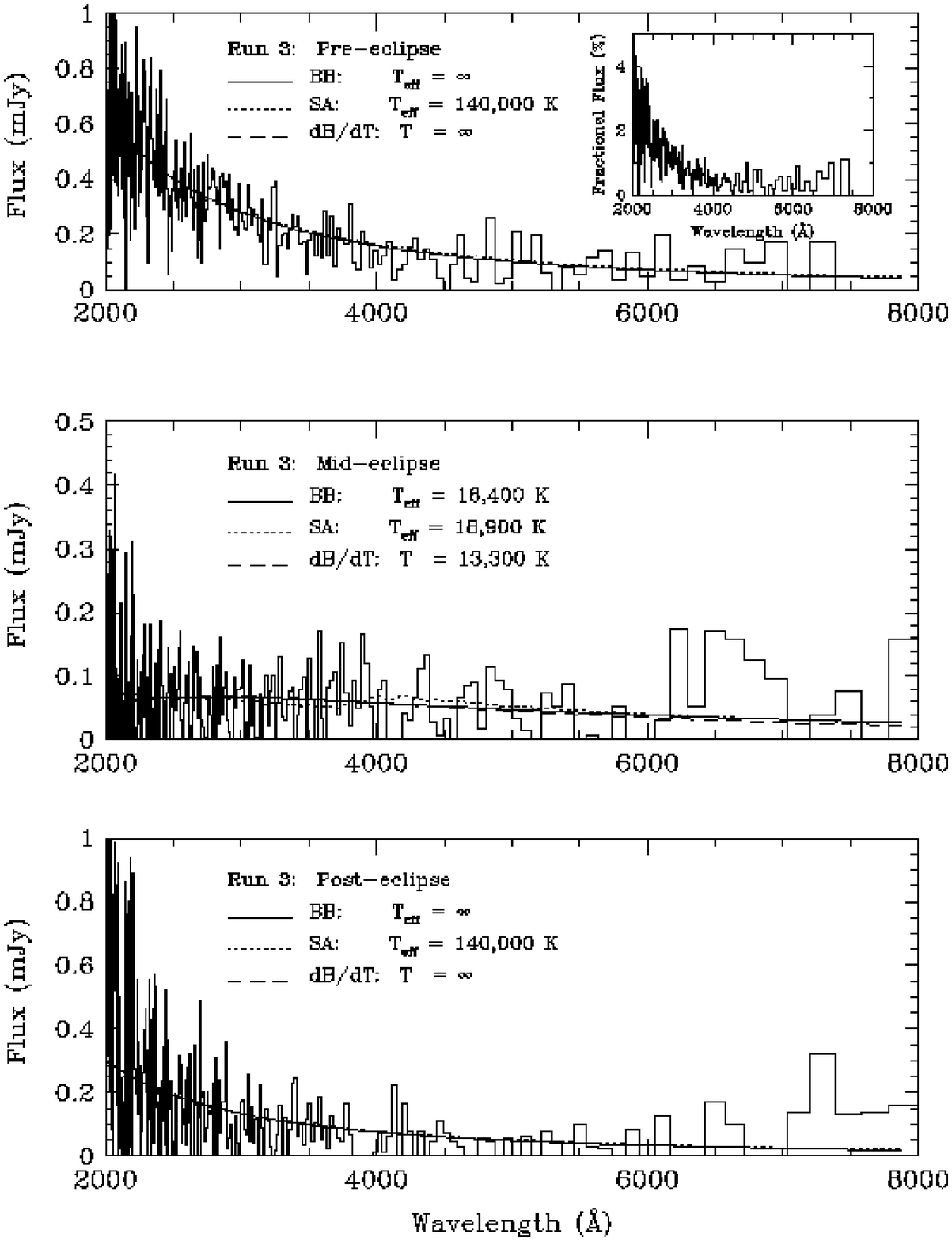]{The pre- (top panel), mid- (middle
panel) and post-eclipse (bottom panel) spectra of 
the 29-s oscillations during Run~3 are shown as the thin histograms. 
The best-fitting BB, SA and $dB/dT$ models are also shown. The inset in
the top panel shows the same oscillation spectrum with the flux 
expressed as a percentage of the flux at a given wavelength in the
overall pre-eclipse spectrum.\label{fig-oscfits1}}

\figcaption[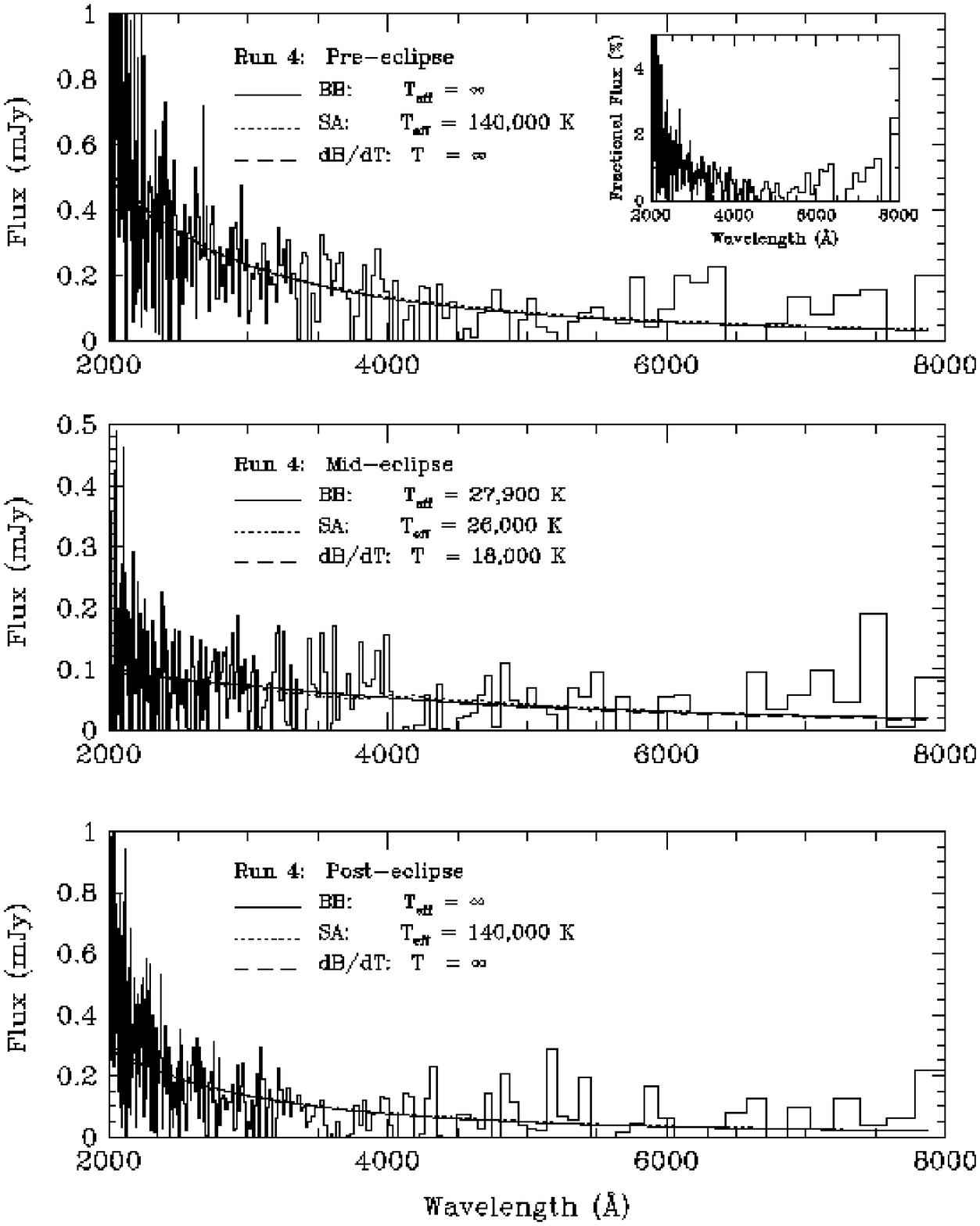]{As Figure~\ref{fig-oscfits1}, but
for the Run~4 data.\label{fig-oscfits2}}

\figcaption[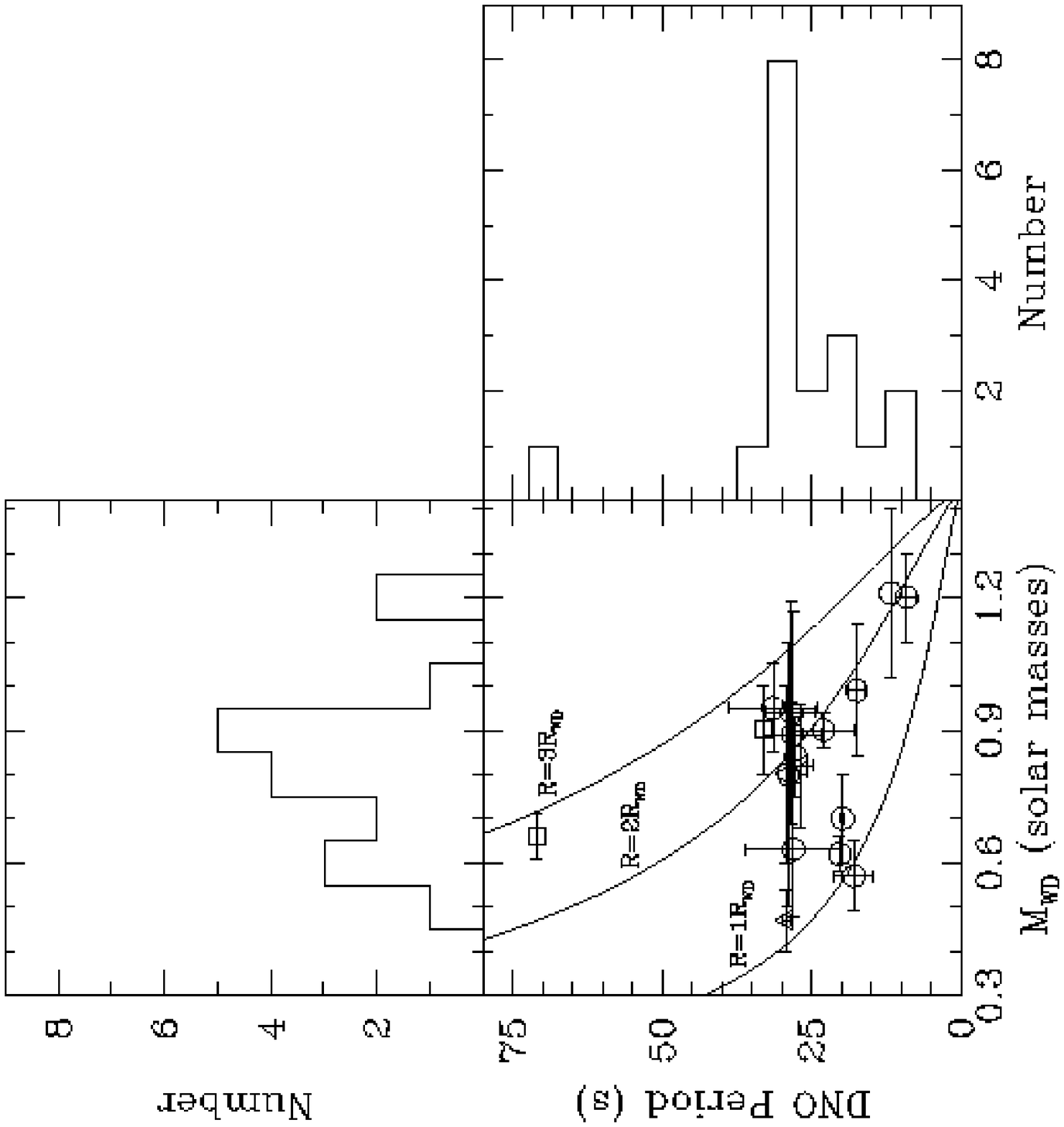]{{\em Bottom left:} The oscillation period
versus WD mass relation for all CVs with mass determinations that are
known to exhibit DN-type coherent oscillations. Open circles are DNe,
open triangles are NLs, open squares are DQ~Her systems (DQ~Her and
AE~Aqr). Vertical error bars correspond to the range of oscillation
periods that have been seen in each system. Oscillation periods and
ranges have been taken from Warner~(1995); white dwarf masses and
errors have been taken from Ritter~(1990), except for UX~UMa for which
the mass estimate of Baptista~{et al.}~(1995) -- $M_{WD} = 0.47 \pm 0.07
M_{\sun}$ -- has been used. The continuous solid lines mark the
rotation periods at the indicated radii in a Keplerian accretion disk 
around a WD of given mass.{\em Bottom right:} Histogram of the DNO period
distribution function. This is just the projection of the data in the 
bottom left panel onto the y-axis.
{\em Top left:} Histogram of the WD
mass period distribution function. This is just the projection of the 
data in the bottom left panel onto the x-axis.\label{fig-alldno}}
\clearpage

\begin{figure}
\figurenum{Figure 1}
\plotone{f1.ps}
\end{figure}

\clearpage

\begin{figure}
\figurenum{Figure 2}
\plotone{f2.ps}
\end{figure}

\clearpage

\begin{figure}
\figurenum{Figure 3}
\plotone{f3.ps}
\end{figure}

\clearpage 

\begin{figure}
\figurenum{Figure 4}
\plotone{f4.ps}
\end{figure}
\clearpage

\clearpage 

\begin{figure}
\figurenum{Figure 5}
\plotone{f5.ps}
\end{figure}
\clearpage

\clearpage 

\begin{figure}
\figurenum{Figure 6}
\plotfiddle{f6.ps}{9in}{0}{100}{100}{-330}{-50}
\end{figure}

\clearpage

\begin{figure}
\figurenum{Figure 7}
\plotone{f7.ps}
\end{figure}
\clearpage

\begin{figure}
\figurenum{Figure 8}
\plotone{f8.ps}
\end{figure}
\clearpage

\clearpage

\begin{figure}
\figurenum{Figure 9}
\plotone{f9.ps}
\end{figure}

\clearpage

\end{document}